%% file: paper.tex
\documentclass[sigplan,screen]{acmart}

\pdfoutput=1

\makeatletter                   %1
\def\mdseries@tt{m}             %1
\makeatother                    %1
\usepackage[plain]{fancyref}
\usepackage[draft=true]{minted} %2
\usepackage{color}
\usepackage{hyperref}           %5
\hypersetup{
    colorlinks=true,
    linkcolor=blue,
    filecolor=red,
    urlcolor=magenta,
    breaklinks=true,            %3
}
\usepackage{breakurl}           %3

\setcopyright{acmcopyright}
\acmPrice{15.00}
\acmDOI{10.1145/3211346.3211348}
\acmYear{2018}
\copyrightyear{2018}
\acmISBN{978-1-4503-5834-7/18/06}
\acmConference[MAPL'18]{2nd ACM SIGPLAN International Workshop on Machine Learning and Programming Languages}{June 18, 2018}{Philadelphia, PA, USA}

\input{macros}

\usepackage{minted}

\usepackage{booktabs}   %% For formal tables:
\usepackage{subcaption} %% For complex figures with subfigures/subcaptions
\usepackage{bcprules}
\usepackage{graphicx}
\usepackage[nounderscore]{syntax}

\graphicspath{ {figures/} }

\begin{document}

%% Title information
\title[Relay]{Relay: A New IR for Machine Learning Frameworks}
% \title{Toward a New IR for Machine Learning Frameworks}
% \subtitle{(Extended Abstract)}
%\title[Short Title]{Full Title}         %% [Short Title] is optional;
                                        %% when present, will be used in
                                        %% header instead of Full Title.
%\titlenote{with title note}             %% \titlenote is optional;
                                        %% can be repeated if necessary;
                                        %% contents suppressed with 'anonymous'
%\subtitle{Subtitle}                     %% \subtitle is optional
%\subtitlenote{with subtitle note}       %% \subtitlenote is optional;
                                        %% can be repeated if necessary;
                                        %% contents suppressed with 'anonymous'

\author{Jared Roesch}
\email{jroesch@cs.uw.edu}

\author{Steven Lyubomirsky}
\email{sslyu@cs.uw.edu}

\author{Logan Weber}
\email{weberlo@cs.uw.edu}

\author{Josh Pollock}
\email{joshpoll@cs.uw.edu}

\author{Marisa Kirisame}
\email{jerry96@cs.uw.edu}

\author{Tianqi Chen}
\email{tqchen@cs.uw.edu}

\author{Zachary Tatlock}
\email{ztatlock@cs.uw.edu}
\affiliation{
  % \position{Position1}
  \department{Paul G. Allen School of \\ Computer Science and Engineering}
  \institution{University of Washington}
  \city{Seattle}
  \state{WA}
  \country{USA}
}

% Thanks Pavel and friends!
% Avoid overlapping text in running head.
\renewcommand{\shortauthors}{Roesch, et al.}

\makeatletter
% This replaces the usual authorship heading with at totally new one
% while overriding a minimal part of the template. Touchable.
\def\addresses{
  \@affiliationfont%
  \centering%
  \begin{tabular}{cccc}%
    \@authorfont Jared Roesch & \@authorfont Steven Lyubomirsky & \@authorfont
    Logan Weber & \@authorfont Josh Pollock \\%
    \nolinkurl{jroesch@cs.uw.edu} &%
    \nolinkurl{sslyu@cs.uw.edu} &%
    \nolinkurl{weberlo@cs.uw.edu} &%
    \nolinkurl{joshpoll@cs.uw.edu} \\ \\%
  \vspace{-0.35in}
  \end{tabular}
  \begin{tabular}{ccc}%
    \@authorfont Marisa Kirisame &
    \@authorfont Tianqi Chen & \@authorfont Zachary Tatlock \\%
    \nolinkurl{jerry96@cs.uw.edu} &%
    \nolinkurl{tqchen@cs.uw.edu} &%
    \nolinkurl{ztatlock@cs.uw.edu} \\ \\%
  \vspace{-0.35in}
  \end{tabular}
  \begin{tabular}{c}%
    Paul G. Allen School of Computer Science and Engineering \\
    University of Washington, Seattle, WA, USA
  \end{tabular}
}
\makeatother

%% Abstract
%% Note: \begin{abstract}...\end{abstract} environment must come
%% before \maketitle command
\begin{abstract}

  Machine learning powers diverse services in industry including search,
  translation, recommendation systems, and security. The scale and
  importance of these models require that they be efficient, expressive, and
  portable across an array of heterogeneous hardware devices. These constraints
  are often at odds; in order to better accommodate them we propose a
  new high-level intermediate representation (IR) called Relay. Relay is being designed
  as a purely-functional, statically-typed language with the goal of balancing efficient
  compilation, expressiveness, and portability. We discuss the goals of Relay and
  highlight its important design constraints. Our prototype is part of the open
  source NNVM compiler framework, which powers Amazon's deep learning framework
  MxNet.
\end{abstract}

%% 2012 ACM Computing Classification System (CSS) concepts
%% Generate at 'http://dl.acm.org/ccs/ccs.cfm'.

%% End of generated code

%% Keywords
%% comma separated list
  %% \keywords are mandatory in final camera-ready submission

%% \maketitle
%% Note: \maketitle command must come after title commands, author
%% commands, abstract environment, Computing Classification System
%% environment and commands, and keywords command.
\maketitle

\input{intro}
\input{background}
\input{lang}
\input{design}
\input{future}
\input{conclusion}

%% Acknowledgments
\begin{acks}
  This work was supported in part by the Center for Applications Driving Architectures (ADA),
  one of six centers of JUMP, a Semiconductor Research Corporation program co-sponsored by DARPA.
  The authors would also like to thank Calvin Loncaric, Pavel Panchekha, Vinod Grover, and
  Eunice Jun for discussion, insightful comments, and feedback on earlier drafts.
\end{acks}

%\clearpage

%% Bibliography
\bibliography{paper}

%% Appendix
%% \appendix
%% \section{Appendix}

%% Text of appendix \ldots

\end{document}

%% file: macros.tex
\usepackage{color}
\usepackage{graphicx}
\usepackage{alltt}
\usepackage{lineno}
\usepackage{proof}
\usepackage{verbatim}
\usepackage{xcolor}

\newcommand{\nnvm}[0]{NNVM }
\newcommand{\tvm}[0]{TVM }
\newcommand{\kwd}[1]{\fcolorbox{gray!30}{gray!30}{#1}}

%% file: intro.tex
\section{Introduction}

Machine learning (ML) has dramatically reshaped computer vision \cite{imagenet_cnn, resnet},
natural language processing, robotics \cite{drl_robot}, and computational biology
and is continuing to gain traction in new areas, including program
synthesis \cite{deep_coder}. Deep learning (DL) in particular has driven progress
in these areas and is now powering diverse services in industry, including search,
translation, recommendation systems, and security. Hardware diversity is growing almost as quickly. DL models are deployed not only in the cloud, but also on a myriad of devices, from off-the-shelf CPUs and GPUs to specialized smartphone chips and IOT edge devices.

The rise of GPUs for DL compute
has made deep learning both possible and scalable for many tasks, but a new generation
of applications with greater compute demands is already upon us. In an attempt to keep up with the ML community's insatiable desire for fast and efficient computation, researchers and industry professionals have introduced a new generation of diverse hardware accelerators and specialized architectures
such as FPGAs and Google's TPU \cite{tpuv1}.

In order to properly schedule, train, and deploy models in a massively distributed,
parallel, and heterogeneous computing environment, scalable systems must be built on
top of specialized hardware to address the challenges of ever-growing available datasets.
Yet even with today's tools, modern systems struggle to satisfy
machine learning's increasing computational demands. Users must
balance interdependent trade-offs at all levels of the DL hardware/software stack. New hardware might mean dramatically
redesigning model architectures, tweaking precision, tuning hyperparameters,
rewriting kernels, and FPGA or ASIC designs \cite{TVMSysML}. Adapting applications and systems
to early accelerators demands substantial rethinking, redesign, and
re-implementation to achieve the best performance. Indeed, multiple projects have already been
undertaken to address this problem, such as Nvidia's Axon, Tensor Comprehensions
\cite{tensor_comprehensions}, and Halide \cite{halide}. As the tools for building
state-of-the-art ML systems grow more complex and varied, it is imperative that
models, hardware, and systems be co-designed and tuned by applying ideas from
synthesis and learning.

\begin{figure}[t]
  \includegraphics[width=\columnwidth]{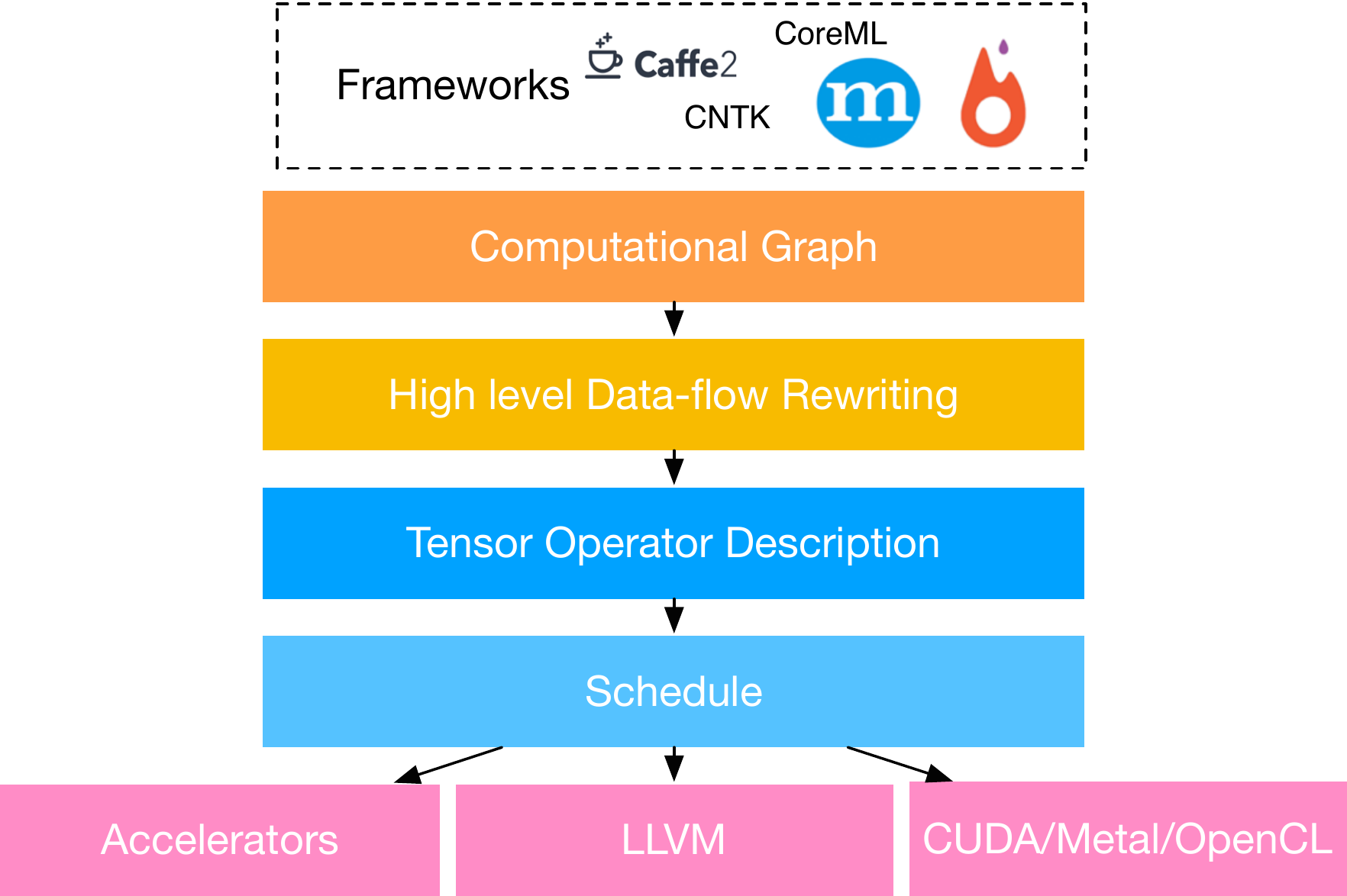}
  \caption{The present TVM stack, as presented in \cite{TVMSysML}. }
  \label{fig:current_tvm}
\end{figure}

\begin{figure}[b]
    \includegraphics[width=\columnwidth]{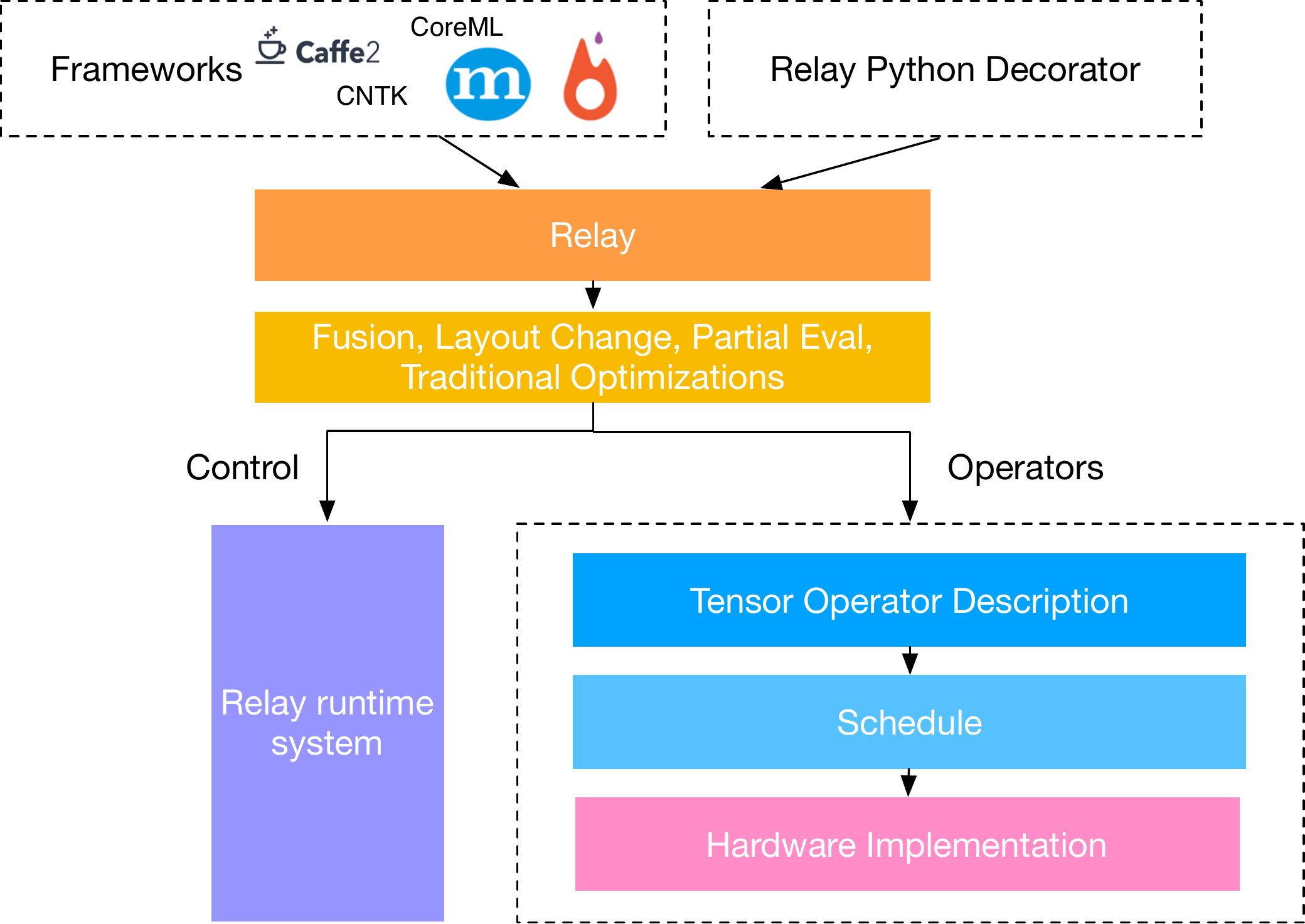}
    \caption{The new TVM stack integrated with Relay.}
    \label{fig:relay_stack}
\end{figure}

\subsection{Existing High-Level Representations}

An end-to-end stack and programming model for heterogeneous
hardware would help satisfy the increasing demands for compute resources and enable
machine learning with specialized accelerators. Research on the lower levels of such
a stack has been realized in the form of the Tensor Virtual Machine (TVM), a hierarchical
multi-tier compiler stack and runtime system for deep learning, depicted in Figure \ref{fig:current_tvm}.

This work is focused on redesigning the top level of the \tvm stack, as depicted in
Figure \ref{fig:relay_stack}. This level, known as NNVM, % did not use macro on purpose, it screwed up spacing before the comma
is based on computation
graphs, the most popular representation of differentiable computation.

Machine learning often relies on computations that are differentiable,
i.e., computations where it is possible to compute a mathematical derivative. In
order to guarantee this property for users' programs, existing frameworks have
limited programs' computational expressivity. Frameworks like TensorFlow
represent differentiable computation using static graphs, which are dataflow graphs with
a fixed topology. These graphs are easy to optimize but require users to
construct programs in a deeply-embedded domain-specific language (eDSL) without
high-level abstractions like functions.

A more expressive style popularized by imperative frameworks like Chainer,
PyTorch, and Gluon allows the construction of graphs with dynamic topologies
that can depend on runtime data and support differentiation of most imperative
computations. This expressivity is convenient for the user but has limited the ability for existing frameworks to optimize user-defined graphs. Moreover, PyTorch's model requires a
Python interpreter, making deployment to new accelerators and FPGAs extremely challenging.

In summary, static graphs are easy to optimize but lack the expressivity
found in higher-level languages; dynamic graphs provide this missing expressivity
but introduce new compilation and execution challenges, especially on heterogeneous
hardware and FPGAs.

\subsection{Relay}
This work proposes Relay, a new high-level intermediate representation (IR)
and language designed to balance efficient compilation, expressiveness, and portability
by combining insights from the approaches of static graphs and dynamic graphs
under the aegis of a functional programming language.

That is, we design Relay not from the perspective of a computation graph but
from that of a programming language for differentiable computation. This PL perspective
will allow Relay to exploit decades of research in functional programming, type
systems, synthesis, rewrite systems, and classical compiler techniques. Our intent in future
work is to demonstrate that these techniques and features will reduce the cost of targeting
new accelerators and enable more optimizations to improve training and inference time, energy
consumption, and space utilization.

This paper presents work in progress towards:

\begin{itemize}
    \item Relay, a new differentiable language for expressing machine learning models.
    \item Higher-order automatic differentiation of Relay programs.
    \item A shape-dependent tensor type system for Relay.
    \item A baseline evaluator, and type-specialized operator compiler built on TVM.
\end{itemize}

%% file: background.tex
\section{Background and Related Work}
\label{background}

Current DL IRs, including NNVM's current
representation, are heavily inspired by dataflow programming and
related computation graph abstractions made popular by previous frameworks.

For example, TensorFlow \cite{tensorflow} is an iteration on previous work at Google, such
as DistBelief \cite{distbelief}. These frameworks have evolved out of dataflow
programming paradigms in which the abstractions are operators with input and
output connections. The semantics provided by these languages have been
sketched in previous work \cite{tf_comp_model}.

TensorFlow employs a dataflow graph of primitive operators extended with restricted control
edges to represent differentiable programs. This representation is sufficient
for many state-of-the-art models and provides an implementation of reverse mode
automatic differentiation~{\cite{ad_survey, tensorflow}}.
TensorFlow can be viewed as a deeply embedded DSL (eDSL), where the result of executing
user's Python script is a computation graph which can then be optimized and
transformed before execution.
Furthermore, because the graph only exposes high-level nodes, it is possible
for the program to be portable to heterogeneous devices, and executing a sub-graph
on a given device requires implementation of only those operators for the device.
 Unfortunately, this programming model has limitations.
Because the topology is fixed before execution, TensorFlow does not lend itself
well to certain applications. As an example, unmodified TensorFlow does not support
building models where the shape of the computation graph is dependent on the input.
While there does exist a library to mitigate this particular problem
(see \cite{tensorflowfold}), this pattern suggests that should new dependencies become
of interest in the future, similar libraries would also have to be written to address each one,
entailing considerable engineering effort.

Dynamic frameworks such as Chainer \cite{chainer_learningsys2015},
PyTorch \cite{pytorch_ad}, Gluon, and TensorFlow eager-mode \cite{tf_eager}
alleviate this problem by moving from the define-then-run model
to the define-by-run model. PyTorch embeds primitives in Python that construct
dynamic dataflow graphs. Control flow is executed in the Python interpreter and
the dataflow is executed by the framework code with is implemented as Python extension.
However when using dynamic frameworks information about control flow is lost, reducing
the ability to optimize them. Additionally, dynamic frameworks need to re-optimize any time
the graph topology changes, costing CPU cycles and the overhead of moving data between
the host and accelerators. This can be solved by transforming the Python code but is effectively
the same as a static framework where Python is the input IR.

Previous work in higher-order differentiation is relevant and has informed
the Relay design. In particular we have drawn inspiration from
various implementations of automatic differentiation \cite{beautiful_diff, ad_survey, haskell_ad, toplas_reverse, wang_reverse, DLS, DDF}.
In particular we are interested in techniques that can compute higher order gradients of higher order programs.

Our work is part of the \tvm stack \cite{TVMSysML}, which is focused on compiling efficient
kernel implementations for deep learning frameworks such as MxNet.

Recent research on the \tvm stack \cite{TVMSysML} has been focused on producing efficient
operators (i.e., dense linear algebra kernels), such as generalized matrix multiplication (GEMM) or
convolutions. This line of research has focused on low-level performance, but demonstrated the need to
tune the high-level computation graph, the operators, and accelerators in tandem to achieve the best performance.
High-level transformations on the input program are especially important for the tensorization problem.
Tensorization is the analogous process to vectorization in existing compilers, and involves the optimizer
decomposing and matching programs to the underlying hardware tensor operations exposed.
This problem is more challenging due to being multi-dimensional, mixed size, and non-finite,
unlike the analogous SIMD primitives.

The \tvm stack is designed to enable a series of fundamental optimizations:

\begin{itemize}
    \item High-level optimizations, such as operator fusion and layout change
    \item Memory reuse at the graph and operators level
    \item Tensorized computations
    \item Latency hiding (traditional hardware provides this abstraction,
          but new accelerators push this burden to the compiler writers)
\end{itemize}

There are multiple related engineering efforts, the primary ones being from Google and Facebook.
Facebook has been building an efficient ML stack composed of many projects
including Tensor Comprehensions \cite{tensor_comprehensions} and Glow \cite{glow}.
Tensor Comprehensions are positioned in a similar space as \tvm, but employs
different techniques, such as using polyhedral compilation rather than algorithmic schedules.
The Glow compiler \cite{glow} is similar to NNVM and intended to
be a compiler for high-level computation graphs. Glow's design is closer to existing
computation graphs, does not appear to be a full language, and is less focused
on full-stack tuning.

TensorFlow's XLA is very similar to the complete TVM stack and is focused on providing
a lower-level intermediate representation for TensorFlow's computation graph. Relay is
designed to replace the user-visible graph with a higher-level abstraction and make it
possible for users to write frameworks like TensorFlow and PyTorch in pure Python.

%% file: lang.tex
\section{Language}

Relay is a statically typed, purely functional, differentiable IR.
Relay is not a low-level IR intended for writing and optimizing high-performance
kernels; rather, it is intended to replace NNVM's computation graph as the input
layer of NNVM. We allow for primitive operators implemented either in external languages
such as C or C++ or in lower-level IRs like TVM or Tensor Comprehensions.
Because Relay is intended as the top layer of the TVM stack \cite{TVMSysML},
we have tight integration with TVM and use it to implement and optimize kernels.

Our intent is for our new IR to serve as a convenient means for researchers
to implement new differentiable programming languages and deep
probabilistic programming languages in the style of Edward and Pyro.

As we discussed in Section~\ref{background}, most popular machine learning
frameworks construct computation graphs that represent the user's program.
Since these graphs are essentially a modified form of an abstract syntax
tree (AST), we consider the transformations and analyses that have been
performed on computation graphs as program transforms and program analyses.
While other DL frameworks also adopt this perspective, their graph-based
approaches have made it difficult to bring the full arsenal of traditional
compiler and programming languages techniques to bear.

Static typing enables direct compilation of models into embedded hardware and
accelerators, which has been demonstrated in prior work done in the TVM
stack~\cite{TVMSysML}.  Having an IR like Relay enables the deployment of
richer dynamic models for applications such as natural language processing.  By
taking this point of view, we can leverage decades of programming language
research to help us express and understand these deep learning models not as a
restricted data flow language, but as a full programming language.

\begin{figure*}[t]
  \begin{minipage}{.5\textwidth}%
    \begin{grammar}
      <Item> ::= <Operator>
      \alt <Definition>

      <Operator> ::= \kwd{operator} <GlobalId> \kwd{:} <Type>

      <Definition> ::= \kwd{def} <GlobalId> (\kwd{(} <LocalId> : <Type> \kwd{)})* -> <Type> \kwd{\{} <Expr> \kwd{\}}

      <Expr> ::= <LocalId>
      \alt <GlobalId>
      \alt $\mathbb{R}$
      \alt \kwd{True}
      \alt \kwd{False}
      \alt <Expr> \kwd{(} (<Expr> (\kwd{,} <Expr>)*)? \kwd{)}
      \alt \kwd{let} <LocalId> \kwd{:} <Type> \kwd{=} <Expr> \kwd{in} <Expr>
      \alt \kwd{(} <Type> \kwd{)} <Expr>
      \alt <Expr> <BinOp> <Expr>
      \alt <UnaryOp> <Expr>
      \alt \kwd{(} (<Expr> (\kwd{,} <Expr>)*)? \kwd{)}
      \alt <Expr> \kwd{[} $\mathbb{N}$ \kwd{]}
      \alt \kwd{[} <Expr> (\kwd{,} <Expr>)* \kwd{]}
      \alt \kwd{if} <Expr> \kwd{then} <Expr> \kwd{else} <Expr>
      \alt \kwd{Zero} <Type>
      \alt \kwd{Grad} <Expr>
      \alt \kwd{Ref} <Expr>
      \alt \kwd{!} <Expr>
      \alt <Expr> \kwd{:=} <Expr>
      \end{grammar}
\end{minipage}%
\begin{minipage}{.5\textwidth}
  \begin{grammar}
    <BinOp> ::=
    \kwd{+}
    \alt \kwd{-}
    \alt \kwd{*}
    \alt \kwd{/}
    \alt \kwd{!=}
    \alt \kwd{=}
    \alt \kwd{\textless}
    \alt \kwd{\textless=}
    \alt \kwd{>}
    \alt \kwd{>=}

    <UnaryOp> ::=
    \kwd{-}
    \alt \kwd{sq}

    <Type> ::=
    <BaseType>
    \alt <Shape>
    \alt \kwd{Tensor} \kwd{(} <Type> \kwd{,} <Type> \kwd{)}
    \alt <Type> \kwd{->} <Type>
    \alt <TypeId>
    \alt \kwd{forall} \kwd{(} <TypeId> \kwd{:} <Kind> \kwd{)} \kwd{,} <Type>
    \alt \kwd{RefType} \kwd{(} <Type> \kwd{)}
    \alt \kwd{(} (<Type> (\kwd{,} <Type>)*)? \kwd{)}

    <BaseType> ::=
    \kwd{IntType} \kwd{(}  $\mathbb{N}$ \kwd{)}
    \alt \kwd{UIntType} \kwd{(}  $\mathbb{N}$ \kwd{)}
    \alt \kwd{FloatType} \kwd{(}  $\mathbb{N}$ \kwd{)}
    \alt \kwd{BoolType}

    <Shape> ::= \kwd{Shape} \kwd{(} ($\mathbb{N}$ (\kwd{,} $\mathbb{N}$)*)? \kwd{)}

    <Kind> ::=
    \kwd{BaseType}
    \alt \kwd{Shape}
    \alt \kwd{Type}

    \end{grammar}
\end{minipage}%
\caption{The BNF Grammar for the Relay langauge. Each case matches a node in our abstract syntax tree.
    References and related operations cannot be included in frontend user code and are only generated by the reverse-mode
    automatic differentiation.}
    \label{ref:lang_def}
\end{figure*}

\subsection{Grammar and Design}

The grammar for the full language can be found in Figure \ref{ref:lang_def}.

Relay is a functional language with closures, recursion, conditionals, operators,
and tensors. Relay's IR has two main design contributions over computation graphs:
the addition of functions and a rich type system that can capture the relationship
of tensor operations.

In order to support higher-order (in the sense of higher-order functions) differentiable
programs, we need to be able to support computing gradients over arbitrary
functions. We accomplish this by introducing a higher-order, higher-order (in
both senses) reverse mode operator \cite{toplas_reverse}. This operator allows
us to compute nth-order derivatives of higher order programs, opening up the
ability to differentiate over arbitrary control structures encoded with
functions.

Inspired in part by DLVM~\cite{dlvm}, a neural network DSL that supports a
CFG-style IR for deep learning programs which introduces a type system for
tensors that is based on constant tensor shape and types, Relay supports a rich
type system that includes dependent typing for tensor shapes, thereby allowing
function type signatures to specify the relationship between arguments (such as
attributes or other tensors) and the resulting tensor shapes.

%% file: design.tex
\section{System Design}

\nnvm currently represents DL programs as static computation graphs
containing operators and input/output data flow. The topology of this graph is fixed,
allowing straightforward compilation to \tvm's graph runtime.

We first constructed a prototype in Python to validate our ideas, and to
experiment with transformations, such as partial evaluation and automatic
differentiation.

Relay is composed of a series of interoperating essential modules:

\begin{itemize}
\item A Python frontend, which translates Python code into Relay's C++ data structures.
\item A module for automatic differentiation of Relay programs.
\item A shape-dependent tensor type system.
\item A simple evaluator for prototyping and debugging.
\item A type-specialized operator compiler built on TVM.
\item An efficient runtime system, which is still in progress.
\end{itemize}

Below, we describe the design and implementation of the modules that have been prototyped and
discuss the in-progress and yet-to-be-implemented components in \ref{future_work}.

\subsection{Frontend}

Relay currently has two interfaces: a textual AST that can be written in Python or C++ and a Python frontend. We intend to add a JSON serialization interface to allow for easy integration with other compilers.

The Python frontend is the intended user-facing interaction
mode for Relay while the other interfaces allow programmatic use of Relay's AST.

The Python interface comprises two pieces: a library and a pair of decorators. The library contains standard DL operators and some Relay-specific functions. The pair of decorators transforms a subset of vanilla Python
code into the Relay textual AST representation and generates a wrapper function
which will execute that code using one of Relay's evaluation mechanisms.

Although the core of Relay is written in C++, we are able to expose the
internals of the system to Python by reusing \tvm's node system, which allows
low-effort interoperability between the two languages. We can expose C++
classes in Python simply by inheriting from a special class and
writing a class stub in Python.

The Python frontend is inspired by many other projects, which use similar
mechanisms to
rewrite Python ASTs, such as Tangent \cite{tangent} \cite{myia}.

Targeting Python has significant advantages since it has become the lingua franca of the DL
community, which is accustomed to Python libraries such as TensorFlow, PyTorch, and Keras. Using Python as a source language also allows users to write and extend Relay in the same language they use to do data processing and deployment.

Figure \ref{fig:dec_ex} demonstrates how to use the decorators, and we will
briefly outline their semantics below.

\begin{figure*}
\begin{minted}{python}
    @relay_model
    def lenet(x: Tensor[Float, (1, 28, 28)]) -> Tensor[Float, 10]:
        conv1 = relay.conv2d(x, num_filter=20, ksize=[1, 5, 5, 1], no_bias=False)
        tanh1 = relay.tanh(conv1)
        pool1 = relay.max_pool(tanh1, ksize=[1, 2, 2, 1], strides=[1, 2, 2, 1])
        conv2 = relay.conv2d(pool1, num_filter=50, ksize=[1, 5, 5, 1], no_bias=False)
        tanh2 = relay.tanh(conv2)
        pool2 = relay.max_pool(tanh2, ksize=[1, 2, 2, 1], strides=[1, 2, 2, 1])
        flatten = relay.flatten_layer(pool2)
        fc1 = relay.linear(flatten, num_hidden=500)
        tanh3 = relay.tanh(fc1)
        return relay.linear(tanh3, num_hidden=10)

    @relay
    def loss(x: Tensor[Float, (1, 28, 28)], y: Tensor[Float, 10]) -> Float:
        return relay.softmax_cross_entropy(lenet(x), y)

    @relay
    def train_lenet(training_data: Tensor[Float, (60000, 1, 28, 28)]) -> Model:
        model = relay.create_model(lenet)
        for x, y in data:
            model_grad = relay.grad(model, loss, (x, y))
            relay.update_model_params(model, model_grad)
        return relay.export_model(model)

    training_data, test_data = relay.datasets.mnist()
    model = train_lenet(training_data)
    print(relay.argmax(model(test_data[0])))
\end{minted}
\caption{An example of the Relay Python decorator, which transforms a decorated
  function into an analogous one in Relay. The defined model is based on
  LeNet \cite{lenet} and is trained and tested on the MNIST dataset.}
\label{fig:dec_ex}
\end{figure*}

Let us preface our description of the decorators by noting that not all of the
functionality in this example is currently implemented and this example instead
represents the design and ideal syntax for our frontend.

To illustrate the decorators, we briefly trace how the frontend transforms the
program in \ref{fig:dec_ex} into Relay.
In this program, three Python functions have been decorated:
\begin{itemize}
  \item \texttt{lenet}: The declaration of the LeNet model \cite{lenet}.
  \item \texttt{loss}: The loss function of the model.
  \item \texttt{train\_lenet}: The training loop.
\end{itemize}
Then we have raw Python code at the bottom for facilitating training and
inference.

Each parameter of the function requires an explicit type annotation, but
the type of local variable assignments can be left out and later be inferred
by the back-end.

Any function call in the \texttt{relay} namespace is converted to an intrinsic
identifier, which must be implemented outside of Relay and registered with the
runtime. TVM is the preferred mechanism for implementing them.

In order to prevent the passing of model parameters to every function that needs
them, we have two separate decorators: \texttt{relay} and
\texttt{relay\_model}. The \texttt{relay} decorator declares a function that
can be run without any hidden state (and thus, no functions that do require hidden
state can be called). The \texttt{relay\_model} decorator declares a function
that cannot be run by default and instead must first be instantiated by a call to
\texttt{relay.create\_model}. When a model is created for a
\texttt{relay\_model}-decorated function, the function's body is searched for any
calls that require hidden parameters; any parameters for these calls are then initialized.
Note that multiple calls to the same function will still generate multiple sets of
hidden parameters. For example, in the \texttt{lenet} function, \texttt{conv1}
and \texttt{conv2} both have their own hidden parameters. Initialization for all
model parameters is currently assumed to be Gaussian with $\mu = 0$ and some
small $\sigma$.

To train the model, we define the \texttt{loss} function in terms of our model
(i.e., \texttt{lenet}), and in our training loop, we use \texttt{relay.grad} to
calculate the gradients of the parameters with respect to the output. Then we
pipe the resulting gradients into \texttt{relay.update\_model\_params} to update
our parameters (this example uses vanilla stochastic gradient descent).
While the Relay IR in general is functional, for convenience, we expose
\texttt{relay.update\_model\_params} as a limited form of mutation.

When training is finished, \texttt{relay.export\_model} returns a callable
version of the trained model that can then be used in raw Python.

\subsection{Automatic Differentiation}
\label{sec:autodiff}

In \cite{toplas_reverse}, the authors demonstrate that reverse-mode automatic
differentiation can be performed in a functional language by using a local
program transformation that introduces references.
Our approach is inspired by their insight and is closely related
to performing forward-mode automatic differentiation using dual numbers.
In the dual number approach, real values are transformed into pairs (called ``dual numbers'')
of the original value and the derivative of the function at that value. All
operations in a function are then lifted to operate over dual numbers.

Instead of pairing each value with its partial derivative, as in the forward mode, we pair each real value
with a reference of type real, denoting the reverse-mode partial derivative. The reverse-mode partial
derivative of a real number is the derivative of that real with respect to the variable representing
the final result of the function \cite{colah}. Additionally, for every reverse-mode AD transformation
we perform, we return a reference to a function from unit to unit, called the ``backpropagator.''
For every real number produced before, the backpropagator is updated to take its partial derivative
and pass it upstream via the references, according to the chain rule. The backpropagator then
calls the old version of itself, thus forming a chain of closures to update every partial derivative.
For a more detailed explanation, see \cite{toplas_reverse}.

We replace each operation over reals with a transformed operation that returns
the original value and a zero-initialized reference, then updates the
backpropagator to clear the gradient reference, propagate the gradient
reference forward, and call the old backpropagator.
To phrase it in more AD-specific terms, the Wengert list is constructed
dynamically as new reals are created. The Wengert list is represented as the
list of closures that created the backpropagator, and the operations to update
the list are bundled with the list.

For every generic operation, including control flow and higher-order functions, we only need to
transform the inner expression and lift the type to accommodate the new expression.
This is identical to what is done in the traditional dual number approach.

Additionally, we extend the syntax with a gradient node (\texttt{Grad expr}). In the gradient node, \texttt{expr}
should be a function from a product of reals to real. The transformed expression is a new function
that calculates the result of the original function bundled with all the partial derivatives.
This node is implemented by transforming the inner AST with our implementation of reverse mode automatic differentiation.
For every real-type argument, we pass the original argument bundled with a new zero-initialized reference to the transformed function.
We call the backpropagator, extract the value in the passed reference, clear the references,
and return the extracted value in a product with the original result.

This transformation requires us to transform every value inside the passed function, so the function
must not contain free variables. (This limitation can always be circumvented by lambda-lifting.)
Given a Relay program without free variables, the transformation always produces a valid Relay
program, meaning it has the closure property. Thus, we have a higher-order reverse mode,
even on programs containing closures.
We take this approach over that in \cite{toplas_reverse} for three reasons:
\vspace{-3pt}
\begin{enumerate}
  \item \textbf{Simplicity}: Pearlmutter and Siskind's approach requires
    reflection on the AST and closure conversion, which means we would need to
    implement reflection, algebraic data types, and closure conversion in our
    own language if we were to follow \cite{toplas_reverse}.
  \item \textbf{Typing}: Additionally, the backpropagators generated in
    \cite{toplas_reverse} have types that depend on the free variables inside
    closures. This means the types of the backpropagators are dynamic, which
    would complicate our type system.
  \item \textbf{Efficiency}: Reflection and traversing the AST are not
    fast. While Pearlmutter and Siskind propose to use partial evaluation to
    remove this overhead, it introduces another layer of complexity.
\end{enumerate}

Currently, we maintain the purity of Relay by only exposing the \texttt{Grad} operation.
User code can never interact with the references that are produced in the above-described process;
the process completely abstracts away the references and returns only the resulting values.
We could also potentially make the code produced by the transformation pure by typing it as lazy functional state
threads (monads), as presented in \cite{lazy_fn_st}.

The implementation of automatic differentiation in Relay comprises 449 lines of C++ out of a total of approximately 10 thousand lines in the C++ backend.

\subsection{A Type System}

Our type system is informed by the authors' previous experience using and implementing dependent type theory. We
have kept the language of types small, inspired by type system designs which use small core languages
\cite{spj_talk, lean_meta}.

Our type system allows shape dependency. That is, it allows types to be polymorphic over shapes which
can appear both in expressions and types. This design allows us to capture important properties
at compile-time, though it sheds the complexity of a traditional dependent type system. Importantly, we
have kinding rules which enforce that shapes and base types are both of a different kind from types of values---
namely tensors, products, and arrow types.

In this paradigm, knowing all values are tensors allows compiler writers to design and implement optimizations over
the AST in a uniform manner. For example, if a user of Relay wants to write an optimization that lifts a
computation up one dimension, they can uniformly add a dimension without needing to handle scalar cases.
This is very useful for optimizations that change dimension (e.g., auto-batching, spatial packing, or layout changes).
We discuss possible extensions to the type system in \ref{future_work}.

The decision to incorporate tensor shape into the type system, rather than to have it as a separate ``analysis,'' allows
shape information to be easily stored and reasoned about at any stage of the optimization pipeline and makes it easier for
users to be explicit about tensor shapes and their desired effect.

% In the present prototype, shapes are
% required to be fixed at constant sizes, but we plan to incorporate a ``shape language'' into the type system to allow for
% more detailed specifications of different operations on tensors.

\begin{figure*}
  \begin{minipage}{.5\textwidth}%
    \infrule[BaseType-T]
       {width \in \mathbb{N}}
       {\Delta \vdash \texttt{IntType}(width) : \texttt{BaseType} \\ \Delta \vdash \texttt{FloatType}(width) : \texttt{BaseType} \\ \Delta \vdash \texttt{UIntType}(width) : \texttt{BaseType} \\ \Delta \vdash \texttt{BoolType} : \texttt{BaseType}}
    \infrule[Shape-T]
      {\\ d_1, d_2, \ldots, d_n \in \mathbb{N}}
      {\Delta \vdash \texttt{Shape}(d_1, d_2, \ldots, d_n) : \texttt{Shape} }
    \infrule[Tensor-T]
      {\\ \Delta \vdash bt : \texttt{BaseType} \andalso \Delta \vdash sh : \texttt{Shape}}
      {\Delta \vdash \texttt{Tensor}(bt, sh) : \texttt{Type} }
    \infrule[Arrow-T]
      {\\ \Delta \vdash T : \texttt{Type} \andalso \Delta \vdash U : \texttt{Type}}
      {\Delta \vdash T \rightarrow U : \texttt{Type} }
  \end{minipage}%
  \begin{minipage}{.5\textwidth}%
    \infrule[Quantifier-T]
      { K \in \{\texttt{Shape}, \texttt{Type}, \texttt{BaseType}\} \\ \Delta, T : K \vdash body : \texttt{Type} }
      {\Delta \vdash \texttt{forall}\ (T : K), \, body : \texttt{Type}}
    \infrule[Product-T]
      {\\ \Delta \vdash T_1 : \texttt{Type} \\ \Delta \vdash T_2 : \texttt{Type} \\ \ldots \\ \Delta \vdash T_n : \texttt{Type} }
      {\Delta \vdash (T_1 \times T_2 \times \cdots \times T_n) : \texttt{Type} }
    \infrule[Ref-T]
      {\\ \Delta \vdash T : \texttt{Type}}
      {\Delta \vdash \texttt{RefType}(T) : \texttt{Type}}
  \end{minipage}%
\caption{Rules for constructing types, indicating kinds. Reference types are only generated internally by reverse-mode automatic differentiation and cannot be given in frontend user code. Also note we will eventually define a more complex AST for shapes.}
\end{figure*}

\begin{figure*}
  \begin{minipage}{.5\textwidth}%
    \infrule[Type-Int-Literal]
      {i \in \mathbb{Z}}
      {\Delta; \Gamma \vdash i : \texttt{Tensor}(\texttt{IntType}(32), \texttt{Shape}())}
    \infrule[Type-Float-Literal]
      {\\ f \in \mathbb{R}}
      {\Delta; \Gamma \vdash f : \texttt{Tensor}(\texttt{FloatType}(32), \texttt{Shape}())}
    \infrule[Type-Bool-Literal]
      {\\ b \in \{\texttt{True}, \texttt{False}\}}
      {\Delta; \Gamma \vdash b : \texttt{Tensor}(\texttt{BoolType}, \texttt{Shape}())}
    \infrule[Type-Tensor-Literal]
      {\\ \Delta \vdash s = \texttt{Shape}(d_1, d_2 \ldots, d_n) \andalso \Delta \vdash b : \texttt{BaseType} \\ \Delta; \Gamma \vdash t_1 : \texttt{Tensor}(b, s) \andalso \Delta; \Gamma \vdash t_2 : \texttt{Tensor}(b, s) \\ \ldots \andalso \Delta; \Gamma \vdash t_m : \texttt{Tensor}(b, s)}
      {\Delta; \Gamma \vdash \lbrack t_1, t_2, \ldots, t_m \rbrack : \texttt{Tensor}(b, \texttt{Shape}(m, d_1, d_2, \ldots, d_n))}
    \infrule[Type-Product]
      {\\ \Delta; \Gamma \vdash p_1 : T_1 \andalso \Delta; \Gamma \vdash p_2 : T_2 \andalso \ldots \andalso \Delta; \Gamma \vdash p_n : T_n}
      {\Delta; \Gamma \vdash (p_1, p_2, \ldots, p_n) : T_1 \times T_2 \times \cdots \times T_n }
    \infrule[Type-Projection]
      {\\ \Delta; \Gamma \vdash p : T_1 \times T_2 \times \cdots \times T_n \\ i \in \lbrack 0, n)}
      {\Delta; \Gamma \vdash p \lbrack i \rbrack : T_i}
   \infrule[Type-Let]
      {\\  \Delta; \Gamma \vdash d : T \andalso \Delta; \Gamma, id : T \vdash b : T^\prime}
      {\Delta; \Gamma \vdash \texttt{let } id = d \texttt{ in } b : T^\prime}
   \infrule[Type-UnaryOp]
      {\\ op \in \{ \texttt{-}, \texttt{sq} \} \andalso \Delta \vdash b : \texttt{BaseType} \andalso \Delta \vdash s : \texttt{Shape} \\ \Delta; \Gamma \vdash t : \texttt{Tensor}(b, s)}
      {\Delta; \Gamma \vdash \texttt{UnaryOp}(op, t): \texttt{Tensor}(b, s)}
   \infrule[Type-Noncomp-BinaryOp]
      {\\ op \in \{\texttt{+}, \texttt{-}, \texttt{*}, \texttt{/} \} \andalso \Delta \vdash b : \texttt{BaseType} \andalso \Delta \vdash s : \texttt{Shape} \\ \Delta; \Gamma \vdash t_1 : \texttt{Tensor}(b, s) \andalso \Delta; \Gamma \vdash t_2 : \texttt{Tensor}(b, s)}
      {\Delta; \Gamma \vdash \texttt{BinaryOp}(op, t_1, t_2): \texttt{Tensor}(b, s)}
  \end{minipage}%
  \begin{minipage}{.5\textwidth}%
    \infrule[Type-Comp-BinaryOp]
      {op \in \{\texttt{=}, \texttt{!=}, \texttt{>}, \texttt{<}, \texttt{>=}, \texttt{<=} \} \\ \Delta \vdash b : \texttt{BaseType} \andalso \Delta \vdash s : \texttt{Shape} \\ \Delta; \Gamma \vdash t_1 : \texttt{Tensor}(b, s) \andalso \Delta; \Gamma \vdash t_2 : \texttt{Tensor}(b, s)}
      {\Delta; \Gamma \vdash \texttt{BinaryOp}(op, t_1, t_2): \texttt{Tensor}(\texttt{BoolType}, s)}
    \infrule[Type-Function-Definition]
      {\\ \Delta; \Gamma, p_1 : T_1, p_2 : T_2, \ldots, p_n : T_n, f : (T_1 \times T_2 \times \cdots \times T_n) \rightarrow T^\prime \\ \vdash body : T^\prime}
      {\Delta; \Gamma \vdash \texttt{def}\ f(p_1 : T_1, p_2 : T_2, \ldots, p_n : T_n) \texttt{ -> } T^\prime, body \\ : (T_1 \times T_2 \times \cdots \times T_n) \rightarrow T^\prime}
    \infrule[Type-Call]
      {\\ \Delta; \Gamma \vdash f : (T_1 \times \cdots \times T_n) \rightarrow T^\prime \\ \Delta; \Gamma \vdash a_1 : T_1 \andalso \Delta; \Gamma \vdash a_2 : T_2 \andalso \ldots \andalso \Delta; \Gamma \vdash a_n : T_n}
      {\Delta; \Gamma \vdash f(a_1, a_2, \ldots, a_n) : T^\prime }
    \infrule[Type-If]
      {\\ \Delta; \Gamma \vdash c : \texttt{Tensor}(\texttt{BoolType}, \texttt{Shape}()) \\ \Delta; \Gamma \vdash b_1 : T \andalso \Delta; \Gamma \vdash b_2 : T}
      {\Delta; \Gamma \vdash \texttt{if}\ c\ \texttt{then}\ b_1\ \texttt{else}\ b_2 : T}
    \infrule[Type-Zero]
      {\\ \Delta \vdash b : \texttt{BaseType} \andalso \Delta \vdash s : \texttt{Shape}}
      {\Delta; \Gamma \vdash \texttt{Zero } \texttt{Tensor}(b, s) : \texttt{Tensor}(b, s)}
    \infrule[Type-Gradient]
      {\\ \Delta; \Gamma \vdash \texttt{autodiff}(e) : T}
      {\Delta; \Gamma \vdash \texttt{Grad}\ e : T}
    \infrule[Type-Ref]
      {\\ \Delta; \Gamma \vdash n : T}
      {\Delta; \Gamma \vdash \texttt{Ref}\ n : \texttt{RefType}(T) }
    \infrule[Type-Val-Ref]
      {\\ \Delta; \Gamma \vdash r : \texttt{RefType}(T)}
      {\Delta; \Gamma \vdash\ !r : T }
    \infrule[Type-Set-Ref]
      {\\ \Delta; \Gamma \vdash r : \texttt{RefType}(T) \andalso \Delta; \Gamma \vdash v : T}
      {\Delta; \Gamma \vdash r := v : () }
\end{minipage}%

    \caption{Rules for deriving types of expressions and definitions. The unit type, $()$, is syntactic sugar
            for a product type with zero members. Note that these type rules assume that all type variables
            in quantifiers have already been concretely instantiated. Additionally, in the rule for
            gradient, ``autodiff'' is the automatic differentiation AST transformation on expression $e$; rather
            than attempt to capture the entire semantics of the transformation in that inference rule,
            we explain the transformation in \ref{sec:autodiff}.}
\end{figure*}

%% file: future.tex
\section{Future Work}
\label{future_work}

Relay generalizes \nnvm's computation graph by moving from a limited dataflow language
to a full programming language. Relay is intended to act as the top layer of
the \tvm stack and serves as its input format. Here we detail near-term future work.

\subsection{Runtime System}

Our current evaluator (an interpreter) is a reference implementation used for differential testing and
experimentation. This evaluator is not sufficient for experimental evaluation, and the main thrust of
our current work is its efficient counter part. An interesting aspect of this evaluator is its
use of TVM as a just-in-time compiler to produce type-specialized tensor operators. The
optimized runtime system, which is intended as the primary way to deploy and execute
Relay programs, is still under heavy development.

Traditional languages have optimized their execution engines'
virtual machines for very specific execution profiles, with long-lived heap
allocations, and relatively small stack values. DL workloads have a
much different execution profile and often do not execute on
traditional CPUs, but rather on special-purpose devices, such as GPUs and
accelerators.

There are many questions about the lifetime of values and how
to handle in-place updates, allocation, reclamation, and more.
The runtime system needs new representations of the
call stack for functions, new allocation patterns
around scopes, and distinct concepts of identity and
allocation.

\subsection{Optimizations}

Relay is designed to provide a whole-program representation of
deep learning programs, allowing us to address problems such as
host slicing \cite{tf_swift}, dynamic networks, change of layout,
latency hiding, and parallel and distributed scheduling. We have designed
Relay with these goals in mind and to help address the critical
optimizations identified in \cite{TVMSysML}.

We envision the ability to add other systems' features as optimization passes
over Relay programs, for example implementing auto-batching from DyNet\cite{dynet},
operator fusion as done in the current NNVM framework, or change of layout for Tensors.
Auto-batching relies on the ability to know about a set of transformations between
unbatched operations and batched operations, inserting the appropriate aggregate
instructions such as summing in the correct places. Given type information
it is possible to extend certain programs with an extra batch dimension,
inserting the appropriate operators to preserve typing and semantics.

\subsection{Software Engineering}

The previous version of Relay supported both a step debugger and the
ability to compile Relay programs to Python for debugging and differential
testing against other machine learning frameworks. We used this to
test automatic differentiation by compiling Relay programs to Python,
using the `autograd' Python library to compute the gradient,
then checking the gradient's results using property-based
testing \cite{quickcheck}.

\subsection{Numerical Accuracy}

ML workloads have proven exceptionally robust to issues of rounding
error~\cite{ml-rounding-error}. Given this tolerance for low-accuracy
arithmetic, we are eager to adapt recent techniques for automatically
rewriting numerical code to improve accuracy at the cost of performance
(e.g., Herbie~\cite{herbie}), to instead trade off accuracy for improved
compute. By adapting tools like Herbie and STOKE~\cite{stoke-fp}
to the context of machine learning inference and training, Relay
will further support developers striving to maximize compute on
platforms built around IEEE-754 floating point arithmetic.  Moving
forward, we hope to further extend these tools and target
specialized numerical representations including mixed width
and fixed point computations; blocked floating point; non-standard,
accelerator-specific numerics; and emerging alternate standards
(e.g., the work on unums and posits~\cite{posits}).

\subsection{Type System Extensions}

One planned type system extension is handling tensors with partially-specified shapes,
that is, shapes where some dimensions are unknown. This is useful for many NLP applications,
where the data may be jagged in one or more dimensions and not representable with a fixed shape.

One other extension is expanding the type system to track individual tensors' data layouts.
This is motivated by the difficulties we have encountered writing change-of-layout optimizations,
which both must infer existing layouts and ensure all uses are transformed. These types of errors
have led to hard-to-debug code that silently produces incorrect results or crashes.
By making these change-of-layout operations explicit, it would be possible to perform
optimizations in that style around automatic boxing and unboxing of values.

% A further extension to the type system would be making the type system more dependent on shapes,
% namely allowing for types that specify tensor shapes that are in some way \textit{calculated}
% from the shapes in parameters or branch based on different parameters (e.g., an axis parameter).
% Having more granular rules in this style would allow the type system to
% express specifications for operations whose output shape can vary and allow the type system
% to easily compute the output shape of a composition of operations without any additional
% manual annotation.

A more significant extension would be an integrated effect system, allowing us to segregate code
manipulating different resources such as random number generators, state, I/O and so on.
This kind of change is more radical and for now is left as an analysis that must be performed
by the compiler.

%% file: conclusion.tex
\section{Conclusion}

We describe an in-progress implementation of Relay: a new IR for efficient
compilation and execution of machine learning models. We are designing Relay
to be the core of the second version of NNVM and to address key challenges
both researchers and engineers face using today's computation graphs. Our initial
prototype implements our vision of how a researcher might write models in Relay,
with the ergonomics of vanilla Python as well as the advantages enjoyed by systems
like PyTorch and TensorFlow. Relay's implementation is still in flux, and we are focused on
exploring topics in section \ref{future_work}. We believe Relay to be an important
part of the TVM stack that will facilitate both current and future research efforts.